\documentclass[aps,prb,showpacs,twocolumn,floatfix,superscriptaddress,longbibliography]{revtex4-1}
\usepackage{graphicx,hyperref,amssymb,amsfonts,amsmath,times,color,soul,epstopdf}
\usepackage{subcaption}  
\usepackage{bbold,bm}
\usepackage{appendix}
\usepackage{subcaption}
\usepackage{adjustbox}

\def\bea{\begin{eqnarray}}
\def\eea{\end{eqnarray}}

\def\frac#1#2{{\textstyle{#1 \over #2}}}

\date{\today}

\newcounter{myparagraphs}

\begin{document}

\title{Critical correlations of Ising and Yang-Lee critical points from Tensor RG}
%Detailed analysis of critical points and critical exponents using Tensor Renormalization Group}

\author{Sankhya Basu}
\affiliation{Physics program and Initiative for the Theoretical Sciences, The Graduate Center, CUNY, New York, NY 10016, USA}
\affiliation{Department of Physics and Astronomy, College of Staten Island, CUNY, Staten Island, NY 10314, USA}
\author{Vadim Oganesyan}
\affiliation{Physics program and Initiative for the Theoretical Sciences, The Graduate Center, CUNY, New York, NY 10016, USA}
\affiliation{Department of Physics and Astronomy, College of Staten Island, CUNY, Staten Island, NY 10314, USA}
\affiliation{Center for Computational Quantum Physics, Flatiron Institute,
162 5th Avenue, New York, New York 10010, USA}

\date{\today}
\begin{abstract}
We examine feasibility of accurate estimations of universal critical data using tensor renormalization group (TRG) algorithm introduced by Levin and Nave \cite{LevinNaveTRG}. Specifically, we compute critical exponents $\gamma, \gamma/\nu, \delta, \eta$ and amplitude ratio $A$ for the magnetic susceptibility from one- and two-point correlation functions for three critical points in two dimensions -- isotropic and anisotropic Ising models and the Yang-Lee critical point at finite imaginary magnetic field. While TRG performs quantitaviely well in all three cases already at smaller bond dimension, $D=16$, the latter two appear to show more rapid improvement in bond dimension, e.g. we are able to reproduce exactly known results to better than one percent at bond dimension $D=24$. 
%in the case appreciable corrections to scaling, making a clear case for uniform convergence in bond dimension apparent in our results.  We are able to reproduce exactly known values to better than 1 percent with modest effort of bond dimension 28. 
We comment on the relationship between these results and earlier results on conformal dimensions, fixed points of tensor RG, and also compare computational costs of tensor renormalization vs. conventional Monte-Carlo sampling.
\end{abstract}

\maketitle

\section{Introduction}
Formulation and subsequent development of stochastic (Monte-Carlo) approaches to the many-body problem is arguably the most significant advancement in computational science to-date. As envisaged in the seminal paper by Metropolis and coauthors \cite{metropolis1,metropolis2}, the method became a defacto general purpose tool, and a must-include topic in even basic coursework.  Although path integral formulation of quantum mechanics enabled significant progress in adapting stochastic techniques to studying quantum problems, the inherent limitations of this approach continue to stimulate research into alternatives.  Perhaps the most successful of these is Density-matrix renormalization group (DMRG) in one-dimensional problems \cite{Schollwock2005,DMRG92} and related tensor extensions to higher dimensions \cite{Verstraete2004}, whose closely related time evolution variants (TEBD, tDMRG) can access short time dynamics -- entanglement growth severely limits the extent of excitation energy and/or time amenable to DMRG techniques in 1D (and, more severely, 2 and 3 dimensional counterparts). DMRG and other variational schemes built around correctly capturing patterns of quantum entanglement remain the go-to method for low dimensional quantum many-body problems. As it turns out some of the basic insights and tools for distilling/managing entanglement are useful for designing or refining other approximation schemes for many-body correlations. An example of that is the tensor renormalization group (TRG) algorithm \cite{LevinNaveTRG} --  a systematic, quantitatively accurate way to implement the celebrated idea of real space coarse graining transformation due to Migdal and Kadanoff \cite{Migdal1,MigdalKadanoff}.  TRG essentially solves the bookkeeping problem inherent in the Migdal-Kadanoff scheme,  relying on singular value decompositions to efficiently compress interactions among block-spins (see Sec. \ref{s:TRGsummary} for further details).  Importantly, the TRG scheme is not restricted to purely statistical problems (characterized by positive Gibbs weights), e.g. in our recent work \cite{Basuetal2021} we discovered symmetry broken steady states in non-unitary dynamics for one dimensional Ising chains, intimately connected with analytic continuation of equilibrium 2D Ising model to complex temperature, and also the complex field Yang-Lee problem \cite{GarciaWeiYL,YangLee52, LeeYang52} we study in this work.
 %adaptated a particular variant of the tensor technique -- the tensor renormation group (TRG) introduced by Levin and Nave\cite{LevinNaveTRG} (see also R.J.Baxter: “Dimers on a Rectangular Lattice” J. Math. Phys. 9, 650 (1968) and \url{http://benasque.org/2021scs/talks_contr/222_Tomotoshi.pdf}) 
 %to help elucidate 
% quantum dynamical phases of 

%This seminal paper established tensor-based renormalization as the feasible method for accurate implementation of conceptually  appealing and historically signifcant block decimation idea (due to Migdal and Kadanoff). The subsequent work can be loosely separated into a few streams -- (i) technical improvements (cite SRG/HOTRG/TNR) aimed at global optimization and better scale separation (removal of short range correllations), (ii) development of scale invariant schemes aimed at better representation of critical states (MERA), and (iii) applications of the method to interesting statmech problems\cite{BerkerTRG,TzuChieh,Meurice,Yuetal2014,Jha_2020,MontangeroMERA}. Our effort here and elsewhere belongs in the latter stream.  
While several improvements have been developed \cite{SRG_PRL,SRG_PRB,HOTRG,EvenblyTNR,LoopTNR}, the plain-vanilla TRG scheme \cite{LevinNaveTRG} may be thought of as analogous to the basic Metropolis algorithm; both of them establishing the essential idea of the method, demonstrating its proof-of-principle usefulness \emph{and} remaining the simplest working implementation of the method sufficient for a quick start on a given problem. % Future improvements and followup work are then motivated by specifics of applications. 
TRG does not suffer from critical slowing down that hampers conventional Monte-Carlo, rather being limited by comparatively more subtle problems. Much of the existing literature on improving TRG focuses on its inability to fully remove short range correlations especially near critical points; however this limitation appears not to hamper the ability to approximate \emph{universal} properties in a convergent fashion, as suggested by our results.

The main thrust of this paper is to establish this point systematically by computing and analysing two-point correlation functions for large separations. Our direct computations readily yield estimates of critical exponents on the order of one percent or even better, consistent with some of the previous works on tensor renormalization interpretting finite size effects using a conformal field theory ansatz\cite{TEFR,EvenblyTNR,LoopTNR}, but well outside of what typical Monte Carlo studies are capable of. Importantly, we benchmark TRG away from well trodden isotropic 2D Ising model and find much clearer evidence for convergence in bond dimension. We also aim to promote a view that for many reasons having to do with ease of implementation, power to access exceedingly large distances \emph{and} potential versatility of being adaptable to study dynamical problems, TRG should be viewed as a viable alternative to Monte-Carlo, and introduced to students early on. With that in mind, the next Section \ref{sec:method} summarizes the basic ingredients of the analysis in a self-contained manner, including a detailed disucssion of errors. Section \ref{sec:results} contains systematic analysis of  critical exponents $\gamma$, $\gamma/\nu$, $\eta$, $\delta=1/\sigma$ as well as the susceptibility amplitude ratio $A$ for isotropic and anisotropic square lattice Ising models, and critical exponents $\sigma$ and $\eta$ at the Yang-Lee critical point.  We conclude with a summary and a few afterthoughts in Sec. \ref{sec:summary}.

\section{Methods}
\label{sec:method}
\subsection{Operator Renormalization with TRG}
\label{s:TRGsummary}

\begin{figure}
\includegraphics[width=\columnwidth]{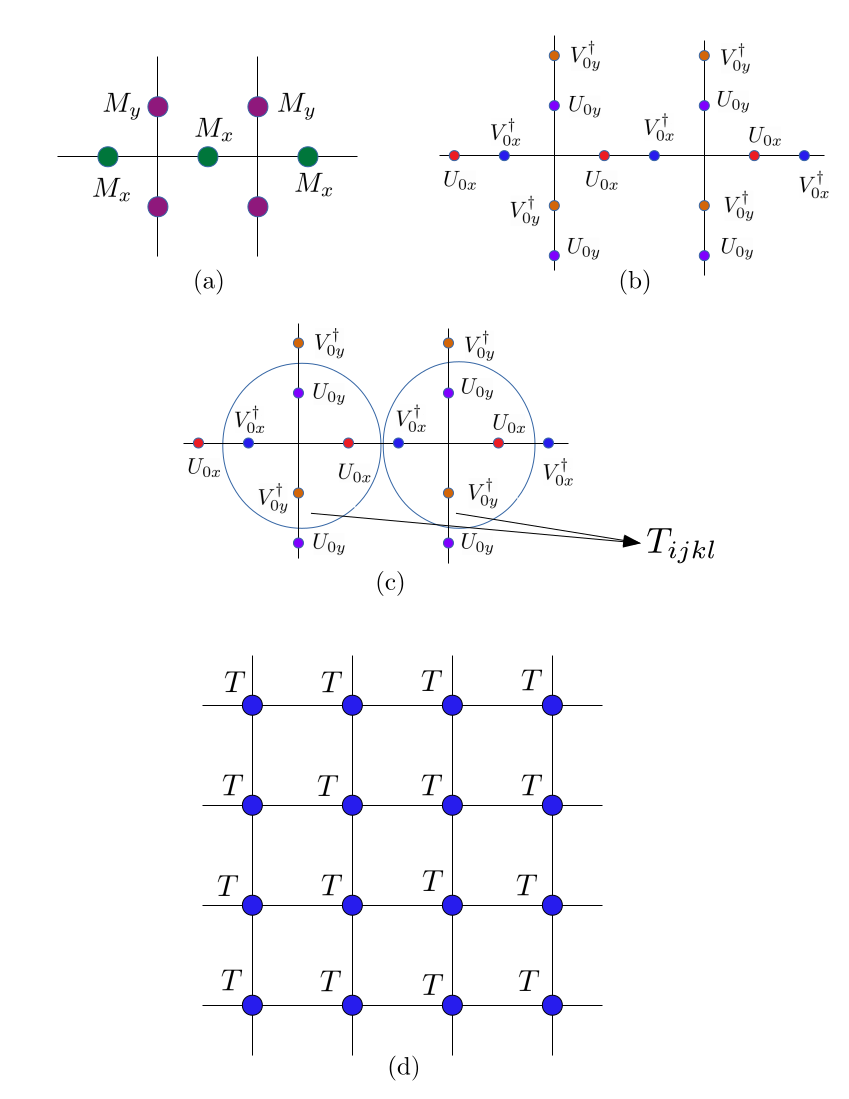}
\caption{Step 1 of TRG. Replacing microscopic degrees of freedom (DOF) living on verices in favor of tensor network and edge DOF. In this implementation each bond's transfer matrix $M$ (panel a) is decomposed using SVD (panel b) with components recombined into  tensors $T$ (panel c).}
\label{f:TRG1}
\end{figure}

The goal of all TRG (and its descendants) is to express partition sums as products of factors ($S_j=\pm 1$)
\begin{align}
&Z_0(\beta, J)=Tr \left[ e^{-\beta H}\right]=\Pi_j z_{0,j}
\\
&Z_1(\beta,J,r)=Tr \left[ S_r e^{-\beta H}\right]=\Pi_j z_{1,j}
\\
&Z_2(\beta,J,r_1,r_2)=Tr \left[S_{r_1}S_{r_2}  e^{-\beta H}\right]=\Pi_j z_{2,j}, \rm{etc}
\end{align}
at inverse temperature $\beta$ (setting $k_B$ to unity), with the standard nearest neighbor spin-spin interactions ($J_{ij}$) and field ($h$) and Hamiltonian
\begin{equation}
H=-\sum_{\langle i,j\rangle} J_{ij} S_i S_j- h \sum_iS_i\; .
\end{equation}
Free energy, magnetization and the two-point correlator can be computed straightforwardly
\begin{align}
-\beta f &\equiv \log Z_0/L^2=\sum_j \log z_{0,j}/L^2,
\\
m&\equiv Z_1/Z_0=\Pi_j z_{1,j}/z_{0,j},
\\
C(r_1,r_2)&\equiv Z_2/Z_0=\Pi_j z_{2,j}/z_{0,j},
\\
C_C(r)&=C(r_1,r_2+r)-m^2,
\end{align}
where we also defined the connected correlator $C_C$.
Importantly, the multiplicative nature of the method results in relatively compact expressions, with number of terms scaling as logarithm of the system size, with typical thermodynamic observables exhibiting rapid convergence in system size, as expected.

\begin{figure*}
\begin{center}
\subcaptionbox{}{\includegraphics[width=0.62\columnwidth]{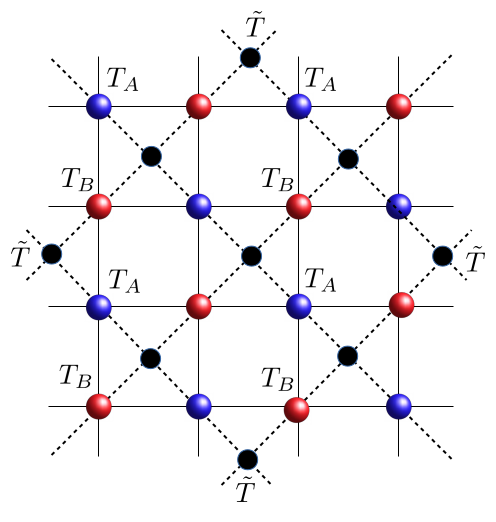}}
\subcaptionbox{}{\includegraphics[width=0.80\columnwidth]{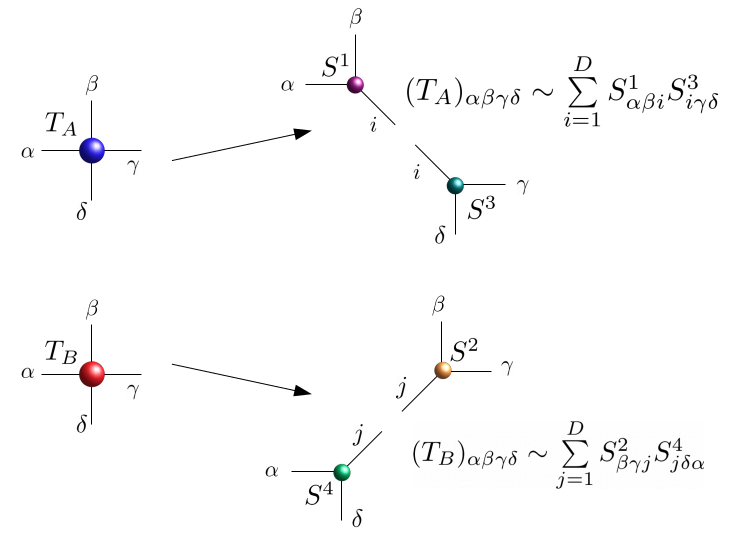}}
\hspace{0.5mm}
\subcaptionbox{}{\includegraphics[width=0.62\columnwidth]{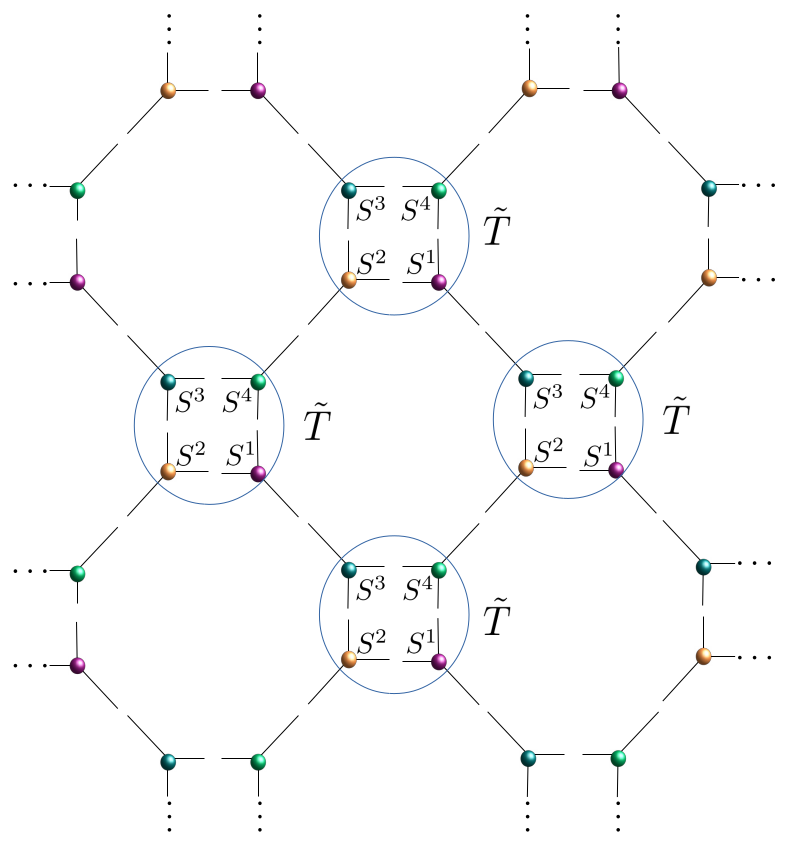}}
\end{center}
\caption{Steps 2 and 3 of TRG. a) tensor networks at the beginning (horizontal lattice with solid interconnected lines) and at the end (lattice rotated by 45$^\circ$ with dotted interconnected lines) of 1 step of renormalization, panel (c) for the formula for renormalized tensors $\tilde{T}$; b) use of SVD to compress initial tensors $T$ and introduce new DOF. For computing correlation functions some of these may be spin operators instead of local Gibbs weight $T$; c) tracing out old DOF to arrive at the new coarse grained tensor network with renormalized tensors $\tilde{T}$ denoted as black filled circles in panel (a).}
\label{f:TRG23}
\end{figure*}

There are three basic steps (see Figs. \ref{f:TRG1} and \ref{f:TRG23}) in going from a microscopic description of the problem (i.e. $H$) to expressions for $z_{n,j}$:  (i) expressing a given problem as a tensor network by tracing out microscopic degrees of freedom (DOF) and introducing new auxiliary DOF; (ii) approximate compression of individual tensors in the network which yields a set of DOF akin to Hubbard-Stratonovich fields optimally chosen to encode local correlations; (iii) tracing out of previous set of auxiliary degrees of freedom (can be done exactly) to arrive at the next tensor network with local geometry equivalent to the previous one, but with fewer, more optimal ``coarse-grained'' DOF.  Steps (ii) and (iii) are iterated until the entire network is coarse grained to just very few tensors that can then be traced over directly. 

None of the steps are unique, e.g. in step (i) the simple Ising model we may be rewritten using plaquette variables (following the seminal paper by Levin and Nave \cite{LevinNaveTRG} or in terms of bond variables, using a singular value decomposition (SVD) of the local transfer matrices $M_x$ (in the $x$--direction)and $M_y$ (in the $y$--direction) \cite{SRG_PRL, SRG_PRB}. The former turns out inferior as it only works at zero field and can lead to slow artifacts in performance on bipartite lattices \cite{BasuUnpub} and we use the latter method to construct our 2D tensor network described below. 
%To do this it is useful to recast the Hamiltonian in the folllowing form
%\begin{equation}
%H = \sum_{\langle i,j\rangle} J_{ij} S_i S_j- \dfrac{h}{n_n} \sum_{\langle i,j\rangle}(S_i + S_j)\
%\end{equation}
%where $n_n$ is the number of nearest neighbors. 
The transfer matrices $M_x$ and $M_y$ are constructed from the Hamiltonian (here we allow for the possibility of anisotropic couplings $J_x\neq J_y$) 
%then 
given by (Fig.\ref{f:TRG1}(a)) 
\begin{align}
M_x = \begin{pmatrix}
e^{\beta(J_x + h/2)} & e^{-\beta J_x} \\
e^{-\beta J_x} & e^{\beta(J_x - h/2)}
\end{pmatrix} \\
M_y = \begin{pmatrix}
e^{\beta (J_y + h/2)} & e^{-\beta J_y} \\
e^{-\beta J_y} & e^{\beta(J_y - \beta h/2)}
\end{pmatrix}
\end{align}
The matrices $U_{0x}, V^{\dagger}_{0x}(U_{0y}, V^{\dagger}_{0y})$ (Fig. \ref{f:TRG1}(b)) are constructed using SVD of the transfer matrix $M_x(M_y)$, which are then subsequently contracted as shown in Fig.\ref{f:TRG1}(c) to derive the 2D tensor network (see Fig. \ref{f:TRG1}(d)).

The ``plain-vanilla'' original TRG algorithm \cite{LevinNaveTRG} uses a simple SVD based ``rewiring'' truncation transformation for step (ii) and an exact trace in step (iii), which is what we stick with in this work, for simplicity and is shown in Fig. \ref{f:TRG23}. One caveat we must note is that the tensors $T_A$ (blue) and tensors $T_B$ (red) in Fig. \ref{f:TRG23}(a) are identical tensors, but labeled differently due to the distinct ways in which one can perform the SVD decomposition on tensor $T$, as shown in Fig. \ref{f:TRG23}(b) to obtain the rank--$3$ tensors $S^1,S^2,S^3$, and $S^4$. The `$S$' tensors are then contracted in the manner shown in Fig. \ref{f:TRG23}(c) integrating out the old DOF to obtain the renormalized tensor $\tilde{T}$. 
In purely statistical problems it is conceptually advantageous to employ the so called Takagi decomposition that produces positive definite renormalized tensors that are amenable to direct block-spin interpretation\cite{Berkeretal, bal2017}

The overall size of the lattice does not appear explicitly in the method, but rather in the number of iterations performed. For a fixed system size of interest the final tensor therefore has elements scaling exponentially in system size, which is not numerically tractable. Instead, a simple normalization scheme is imposed to rescale the renormalized tensor $\tilde{T}$ after each iteration to line it up with the previous $T$, with rescaling factors $z_{m,j}$ collected in this process, except for the last one, which comes from tracing out the renormalized final tensor network (typically made up of very few tensors).

TRG can achieve an exceptionally high degree of accuracy at small or moderate computational cost, especially in two-dimensional problems. This is generally understood as a consequence of individual coarsegraining steps being local in space and well controlled through application of SVD \emph{and} rapid convergence to relatively simple fixed point tensors, e.g. in the paramagnetic phase, and so the individual errors in the method are small both at the beginning of the flow and also late in renormalization. As the only explicitly approximate step is the compression step (ii) using SVD we shall refer to the numerical errors of the algorithm broadly as ``finite bond dimension'' errors. For some quantities these errors are heavily dominated by initial steps in renormalization, due to volume factors (the size of the lattice shrinks exponentially fast under renormalization), e.g. the expression for the free energy above can be rewritten as $f\to \sum_{j'=0}^j f_{j'}/2^{j'}$ with $f_j\lesssim 1$ to exhibit this predominance of short distance contributions.  As we are more interested in universal properties, such short distance dominated errors are not especially worrisome or interesting. They do appear, most prominently, as non-universal bond dimension dependent shifts of the critical temperature. The pattern of short distance errors appears erratic leading to these errors, while small, also being of random sign and poorly convergent.

The more delicate, and in a certain sense fundamental, issue is whether any given RG scheme reaches the correct fixed point tensor. It is empirically known that the original TRG scheme does so sufficiently far away from the critical point, correctly capturing paramagnetic (PM) and ferromagnetic (FM) phases of the Ising model corresponding to infinite and zero temperature fixed points, respectively. There appears a range of temperatures (that presumably shrinks with bond dimension) near the critical temperature $T_c$ where the fixed point tensors of the transformation are non-critical so-called corner double line (CDL) tensors, distinct from the PM/FM ones\cite{TEFR,EvenblyTNR} (this is also apparently true of some of the other schemes, e.g. HOTRG\cite{lyu2021}).  It is unclear, based on published previous work, how these spurious fixed points of tensor renormalization depend on the initial choice of a mapping to a tensor network, lattice symmetry and especially presence (and magnitude) of small symmetry breaking fields. 

There have been several improvements to TRG to address (remove) this artifact through better local coarse graining steps and also work to implement global optimization of the coarse graining process.  The detailed nature of these unphysical fixed points, their significance to long distance correlations \emph{and} their robustness against weak symmetry breaking fields have only been partially elucidated.  Our analysis of \emph{universal} content of critical points in this work helps fill that gap, especially in regard to the last aspect, i.e. introduction of very small symmetry breaking field.  We expect (and find) that long distance properties have finite bond dimension errors that are of a fixed sign as they originate (presumably) from how well a finite dimensional TRG fixed point tensor approximates the true critical fixed point tensor that is infinite dimensional. Unlike variational tensor based approaches using matrix product states (MPS), where finite bond dimension %relatively straightforwardly translates into 
implies finite correlation length rounding of the phase transition, it is not clear (to us) whether/how similar length generation takes place in the present context.  Empirically, based on the data we present below (see Figs. \ref{f:allisingplots} and \ref{f:YLplots}) for the two-point correlation function, such length scale, if present, is either astronomically large or does not appear in simple two-point correlations.
\\
\\
Lastly, we survey prior efforts and methodology to study critical exponents  using tensor renormalization. By far the most common approach is to assume that tensor renormalization correctly reproduces the expected conformal field theory operator structure and use the goarsegrained tensor of a finite lattice to identify conformal dimensions of operators from finite size correction to the partition function by following Cardy's ansatz, i.e. without explicitly computing correlation functions\cite{TEFR,EvenblyTNR,LoopTNR}. Alternately, in one of the earliest attempts, Hinczewski and Berker\cite{Berkeretal} estimated the thermal eigenvalue of the RG flow by directly computing the tangent map of TRG to find its Lyapunov spectrum. This beautiful application of basic renormalization group seemed to work well for small bond dimension but then authors encountered strange flow instabilities and had to resort to the more brute force method of differentiating the free energy and examining the resultant specific heat and the associated exponent $\alpha$ (see also a much more recent effort along these lines by Lyu and collaborators\cite{lyu2021}). The specific heat singularity was also used to locate $T_c$ in other works\cite{Berkeretal}.  Off-critical correlation functions were computed away from $T_c$ by Nakamoto and Takeda\cite{TakedaTRGcorrelation} and used to estimate  exponents $\beta$ and $\nu$. Importantly, the calculation of the two-point function simplifies to a concatenation of two one point function flows for simple exponential families of spatial distances, e.g. $r=(0, 2^j), (2^j,0), (2^j,2^j)$ on the square lattice with steps up to $j$ tracking the renormalization of bulk tensors and that of the spin, then fusing them into a new composite operator and renormalizing that for larger $j$. Such exponentially separated set of points is also best for studying powerlaws using log-log plots and so we exploit this double advantage in our analysis below. In summary, while the CFT method is by far the most powerful and elegant of these, we opted for explicit renormalization of one- and two-point operators to examine observables directly and study criticality with minimal set of assumptions, anchored in direct observation of scale invariant behavior.

\subsection{Locating the Critical Point and Estimating Exponents}
%We start by recalling the general protocol for estimating the universal content of critical points. 
The critical point is a singularity in system's macroscopic properties. It is a low dimensional manifold in the infinite dimensional space of all parameters, corresponding to the so called relevant operators. For the Ising transition there are only two -- temperature and field, so we write $m(T_c-T,h)\sim (T_c-T)^\beta, \sim \rm{sign}(h) |h|^\sigma$ for $h\to 0^+$ and $T=T_c$, respectively.  The susceptibility $\chi \equiv \lim _{h \to 0}\dfrac{\partial m}{\partial h}$, follows $\chi_> (T>T_c) \to A_> (T-T_c)^{-\gamma}$ (and similarly for $T<T_C$) and $\chi_{\text{max}}(T=T_c)\sim L^{\gamma/\nu}$, where $L$ is the linear system size. The two point correlator at the critical point $T=T_c$ follows $C(r_1-r_2)\sim |r_1-r_2|^{-\eta}$ (in two-dimensions, also in the generic case; see the Yang-Lee case below for an exceptional case). %, thus defining exponent $\eta$.  
Previous studies have examined exponents $\beta$, $\nu$ and $\alpha$ by directly computing the observables (as we do here)\cite{Berkeretal, LevinNaveTRG}, this paper will tackle $\eta, \gamma, \sigma$, as well as the  amplitude ratio $A=A_>/A_<$ which is also universal\cite{delfino1998universal}. % Other studies have also examined the critical exponent $\delta$ for the 2D XY model \cite{Yuetal2014,Jha_2020} using magnetic suscpetibility to locate the critical point. the  
Since the numerically found exponents will turn out close to their known values we will not be using hyperscaling to further test the consistency of the results, which otherwise would have been a useful check.

We shall use the divergence of the susceptibilty to locate the crtical temperature $T_c(D)$, which shifts compared to its true location due to finite bond-dimension $D$ approximation employed. 
Locating the critical point is the required first step in any systematic study of the critical point. Past work showing breakdown of TRG seems not to have focused on this point carefully\cite{EvenblyTNR} or used other critical ``markers'', e.g. the specific heat singularity\cite{Berkeretal} (which  may be adequate for the case of the 2D Ising model but not generally) or the divergence of correlation length in the paramagnetic phase \cite{TakedaTRGcorrelation}
. The divergence in the susceptibility is both generic for symmetry breaking transitions and tends to be the most prominent, and hence is the natural choice for locating the critical point in a finite sized sample. We also note that our $T_c$ estimates are similar or somewhat better to those obtained by other methods\cite{Berkeretal,TakedaTRGcorrelation} 
%already 
despite comparatively more modest bond dimension values, which further supports our approach to locating the critical point. Recalling that  iteration index in TRG is the logarithm of the sample size (for any bond dimension) we expect the susceptibility to diverge exponentially with TRG iteration at $T=T_c$, hence, we can estimate the relevant combination of critical exponents, $\gamma/\nu$, from the rate of this exponential growth.%, i.e. its leading Lyapunov exponent. %(note -- logarithmic specific heat singularity studied in the past\cite{berker} is not amenable to such analysis).  
We also use onset of spontaneous magnetization (properly computed in the presence of small field -- more below) as an alternative estimate of transition point and observe that the two methods correlate well, but not perfectly, of course. Lastly, we expect and observe scale invariant decay in the 2-point correlation function at criticality, up to distances that depend sensitively on residual relevant operators, e.g. imperfect tuning of $T-T_c$ or small applied field.  Thus, to summarize, our convidence in locating $T_c(D)$ rests on consensus among three computed properties -- agreement between magnetization onset (plots not shown), peak in $\chi(T)$ with powerlaw divergence of critical susceptibility $\chi_{\text{max}}(T=T_c)\sim L^{\gamma/\nu}$, \emph{and} powerlaw decay of the critical two-point function with separation $C(r)\propto r^{-\eta}$, listed here in the order of increasing sensitivity, e.g. deviations from powerlaws appear on shorter scales in the correlation function than susceptibility\footnote{Note, %while we expect our estimates of $T_C$ to correlate well with ones based on 
Monte-Carlo simulations  employ magnetization (Binder) cumulants to locate the critical point -- these are not accessible in the simplest TRG formulation employed here.  However, typical lattice sizes amenable to Monte-Carlo sampling are comparatively small and which makes our direct method less reliable, presumably. }
. 
Importantly, all this analysis is conducted in the presence of a small symmetry breaking field, typically $h\lesssim 10^{-12}$. We have heuristic evidence that the presence of this field improves convergence of TRG without disrupting powerlaws out to distances  $\gtrsim 2^{12}$.

Having established the location of the critical point (for a particular bond dimension), we can readily obtain estimates of the critical exponent $\gamma$, the amplitude ratio $A\equiv A_>/A_<$, and the combination of critical exponents $\gamma/\nu$, from the finite size cutoff of susceptibility precisely at $T=T_c$. 
We also compute the two-point correlation function at large separation, which is a particularly direct measure of criticality, from which we extract the $\eta$ exponent. Finally, we extract $\delta$ from the nonlinear magnetization at $T=T_c$, to round out the roster of standard exponents that have not been computed previously using TRG (to the best of our knowledge).

\begin{table}
\includegraphics[width=\columnwidth]{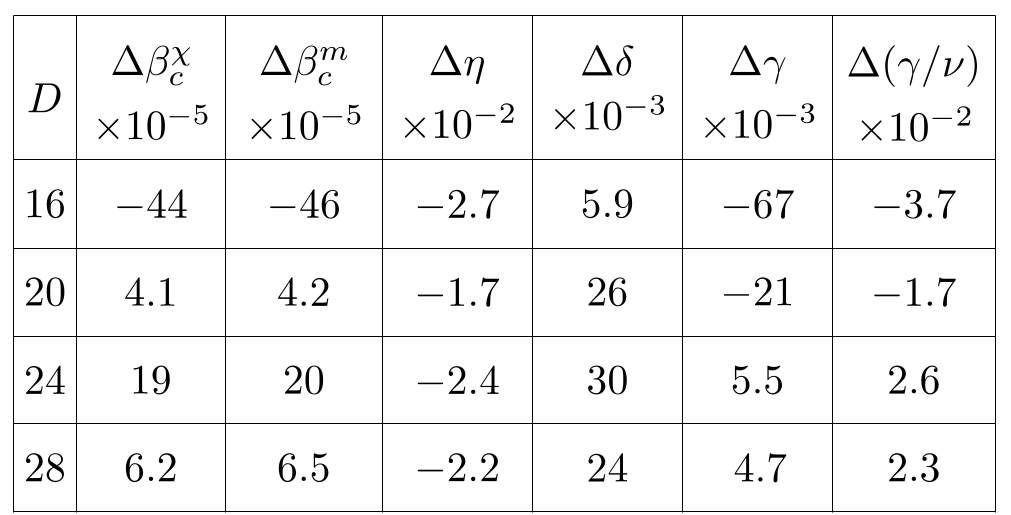}
\caption{Summary of universal and non-universal data on the isotropic Ising critical point, see text for definition of absolute error $\Delta$}
\label{t:isotropictable}
\end{table}
\begin{table}
\includegraphics[width=\columnwidth]{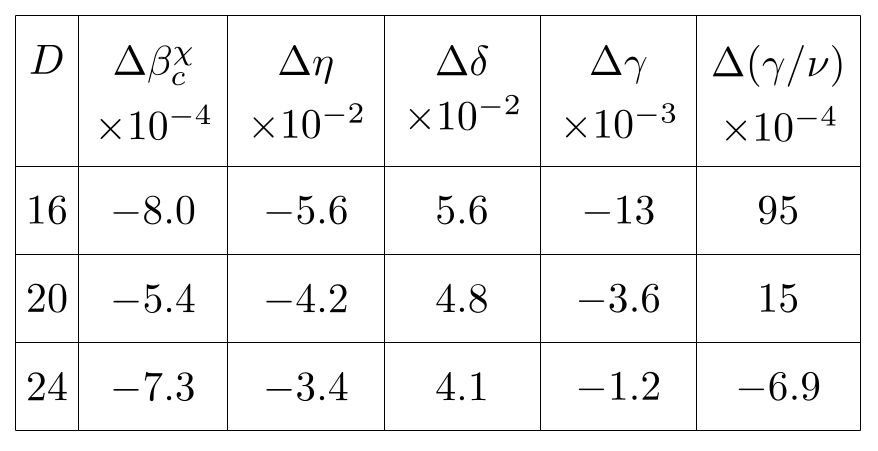}
\caption{Summary of universal and non-universal data on the anisotropy 10 Ising critical point.}
\label{t:anisotropictable}
\end{table}

\begin{figure}
\includegraphics[width=\columnwidth]{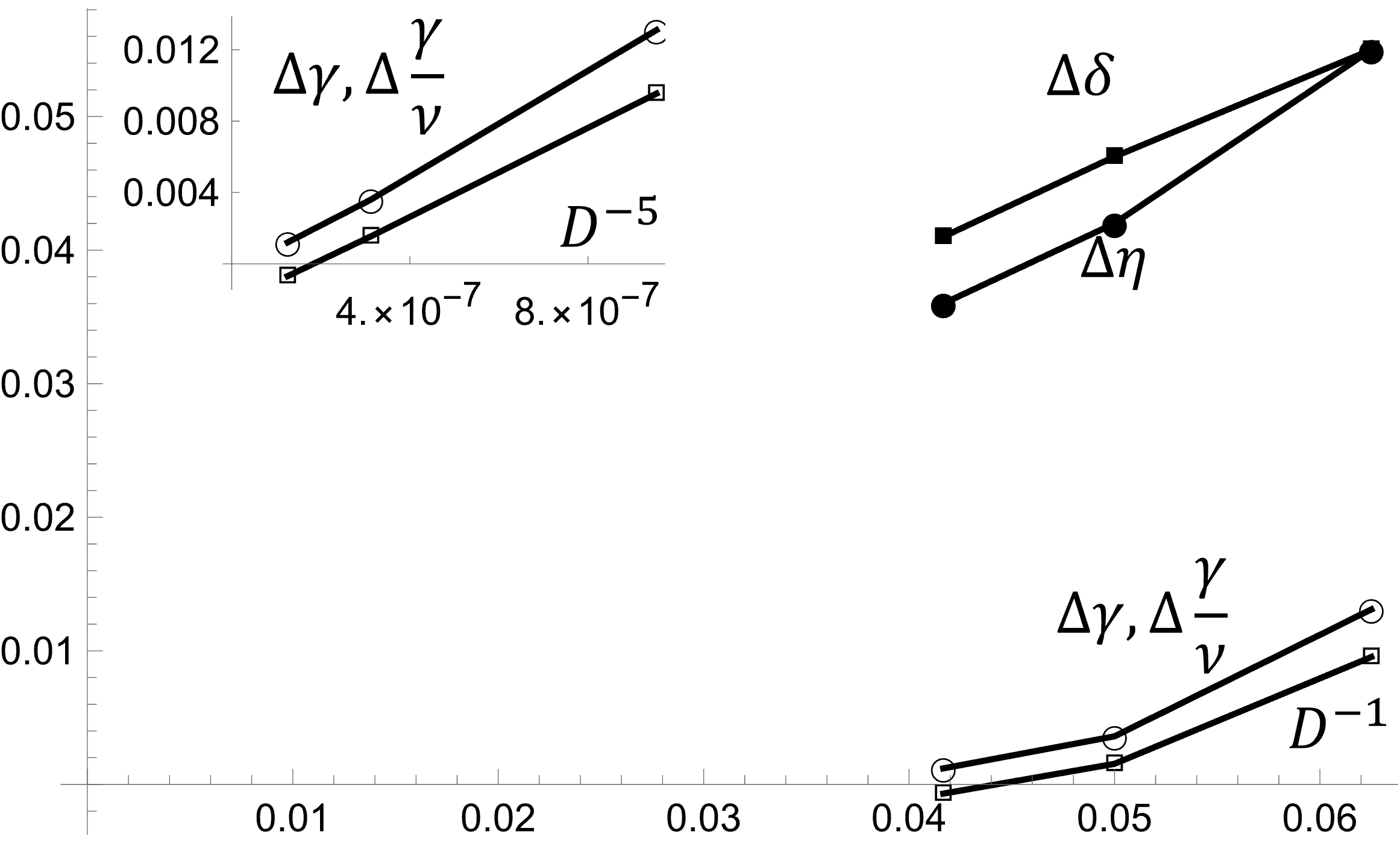}
\caption{Convergence trend for exponents $\delta$, $\eta$ (closed symbols) and $\gamma$, $\gamma/\nu$ (open symbols) for the anisotropy 10 model. (Inset) Visual test of the empirical $D^{-5}$ behavior.}
\label{f:anisconvtrend}
\end{figure}
 
\subsection{Errors}
%Before turning to specific implementations of this strategy we comment on errors. 
There are two sources of technical error in this analysis -- (i) finite resolution in the values of magnetic field $h$ and temperature $T$ (relevant operators), which may be refined as needed and (ii) finite bond dimension, which are not thoroughly understood and commonly dealt with by switching to more elaborate (and accurate) coarsegraining schemes. As all of these are systematic errors, one cannot rely on added runs to improve results, e.g. as with Monte-Carlo simulations. The latter, of course suffer from their own problems inherent in sampling large lattices, including but not restricted to critical slowing down, all of which are entirely absent here. While finite bond dimension can be related to finite length rounding of critical correlations in variational MPS based approaches to classical criticality\cite{burgelman2022contrasting}, the exisence of such finite length rounding is far from obvious in TRG and other multiplicative renormalization approaches. Thus it is somewhat unclear (to us) how to relate technical errors that are clearly present in the computation to the physical errors (if any) in conclusions drawn. We shall return to this point in Sec. \ref{sec:summary} after presenting the results in the next Section (see also discussion above, in Sec. \ref{s:TRGsummary}).

%We strive to remove the first type of an error, typically using the quality of powerlaws themselves to gauge whether sufficient resolution was achieved. We found that application of small fields is absolutely crucial to obtain sensible results near/below the transition, e.g. for the two-point correlation function to correctly factorize at long distances (presumably due to residual cat-like entaglement supported by TRG?). Yet, since field is a relevant perturbation it must be small enough to avoid rounding the singularity in the susceptibility, values on the order $10^{-12}$ appeared sufficient.   Locating the critical temperature is done iteratively, first by scanning for the peak in $\chi (T)$ and then examining two-point correlators obtained at the peak temperature and two adjacent temperatures -- at large separation the correlation function clearly displays the influence of the residual relevant operator which allows, in principle, further refinement of subdivision of the temperature axis. (ii)

\begin{figure*}
\begin{center}
\vspace{-3mm}
\begin{minipage}[t]{0.05\columnwidth}
\subcaption{}
\end{minipage}
\begin{adjustbox}{minipage=0.88\columnwidth, valign=t}
\includegraphics[width=\columnwidth]{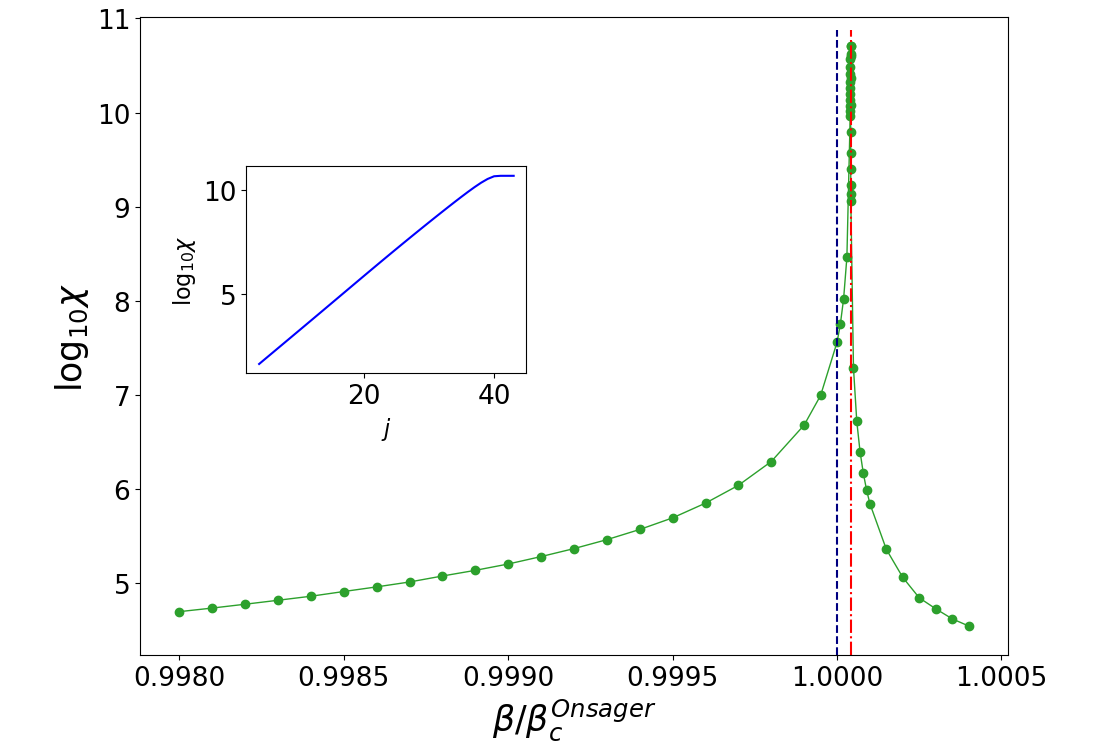}
\end{adjustbox}
\quad
\begin{minipage}[t]{0.05\columnwidth}
\subcaption{}
\end{minipage}
\begin{adjustbox}{minipage=0.88\columnwidth, valign=t}
\includegraphics[width=\columnwidth]{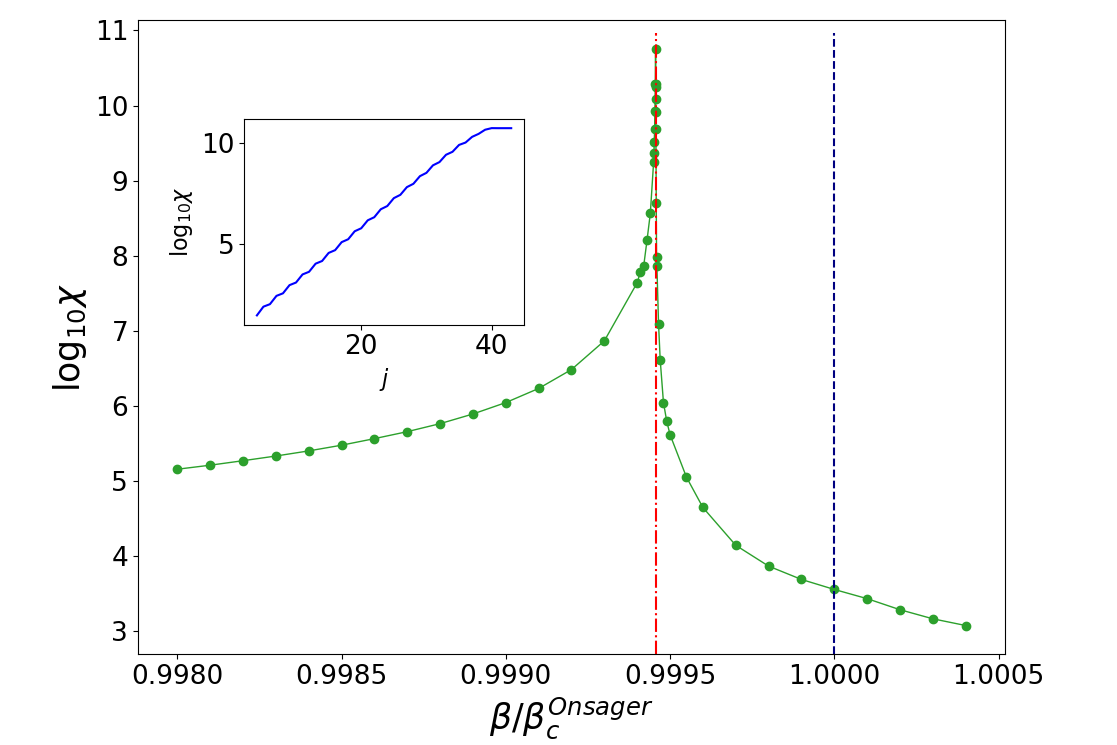}
\end{adjustbox}

\begin{minipage}[t]{0.05\columnwidth}
\subcaption{}
\end{minipage}
\begin{adjustbox}{minipage=0.88\columnwidth, valign=t}
\includegraphics[width=\columnwidth]{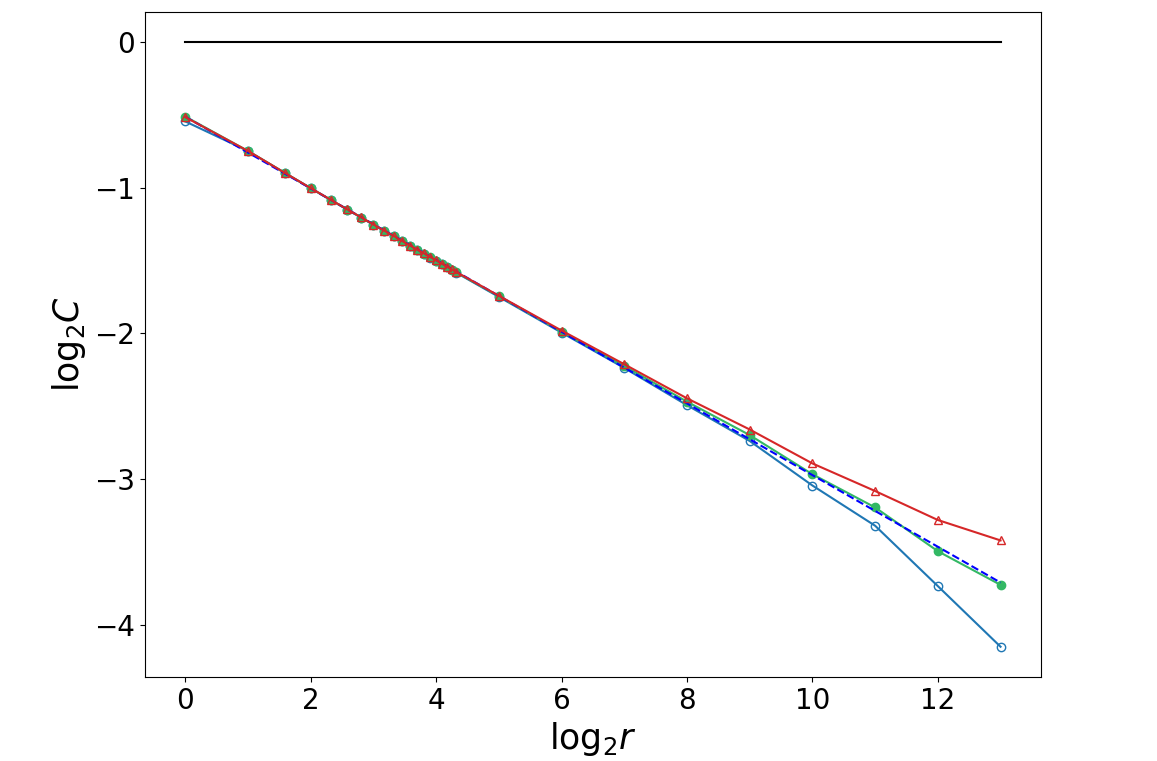}
\end{adjustbox}
\quad
\begin{minipage}[t]{0.05\columnwidth}
\subcaption{}
\end{minipage}
\begin{adjustbox}{minipage=0.88\columnwidth, valign=t}
\includegraphics[width=\columnwidth]{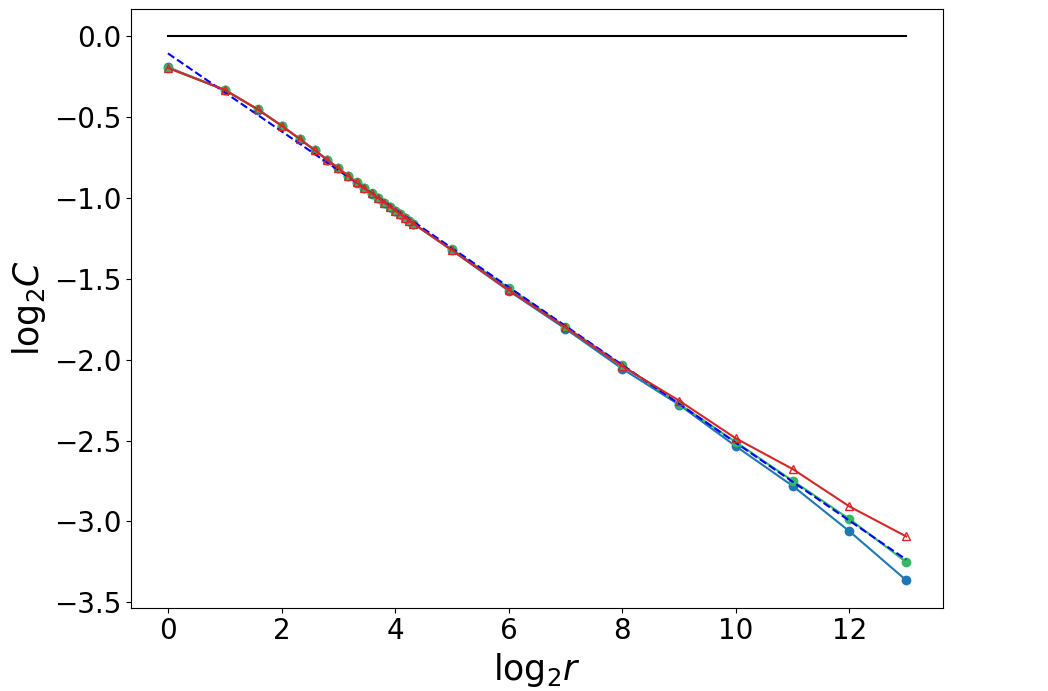}
\end{adjustbox}

\begin{minipage}[t]{0.05\columnwidth}
\subcaption{}
\end{minipage}
\begin{adjustbox}{minipage=0.88\columnwidth, valign=t}
\includegraphics[width=\columnwidth]{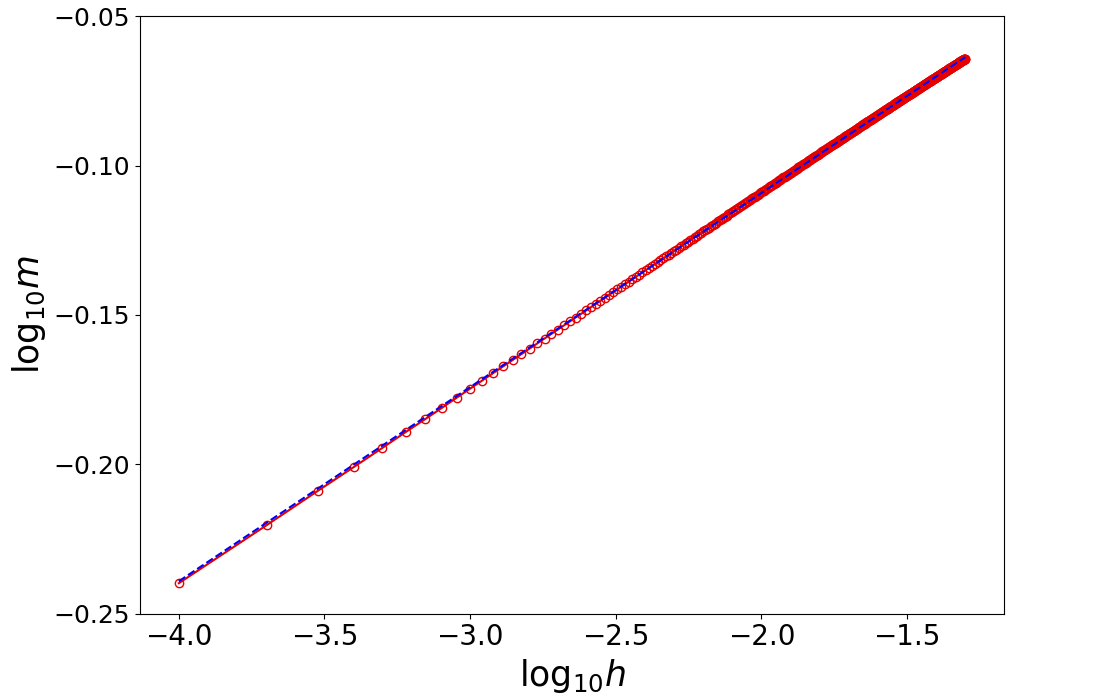}
\end{adjustbox}
\quad
\begin{minipage}[t]{0.05\columnwidth}
\subcaption{}
\end{minipage}
\begin{adjustbox}{minipage=0.88\columnwidth, valign=t}
\includegraphics[width=\columnwidth]{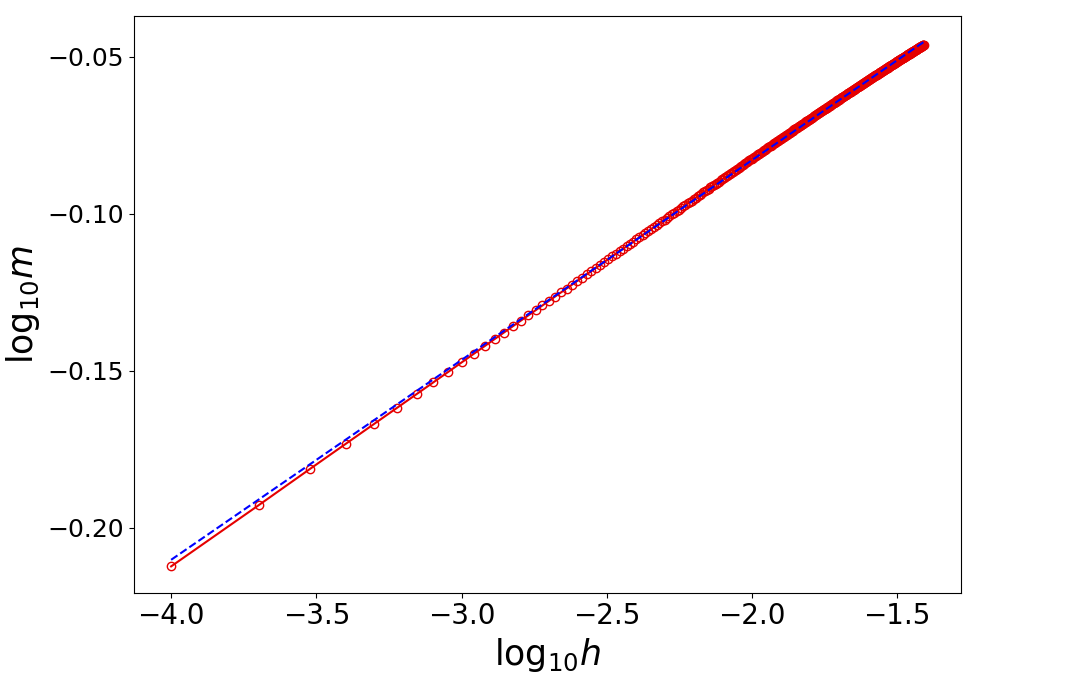}
\end{adjustbox}

\begin{minipage}[t]{0.05\columnwidth}
\subcaption{}
\end{minipage}
\begin{adjustbox}{minipage=0.88\columnwidth, valign=t}
\includegraphics[width=\columnwidth]{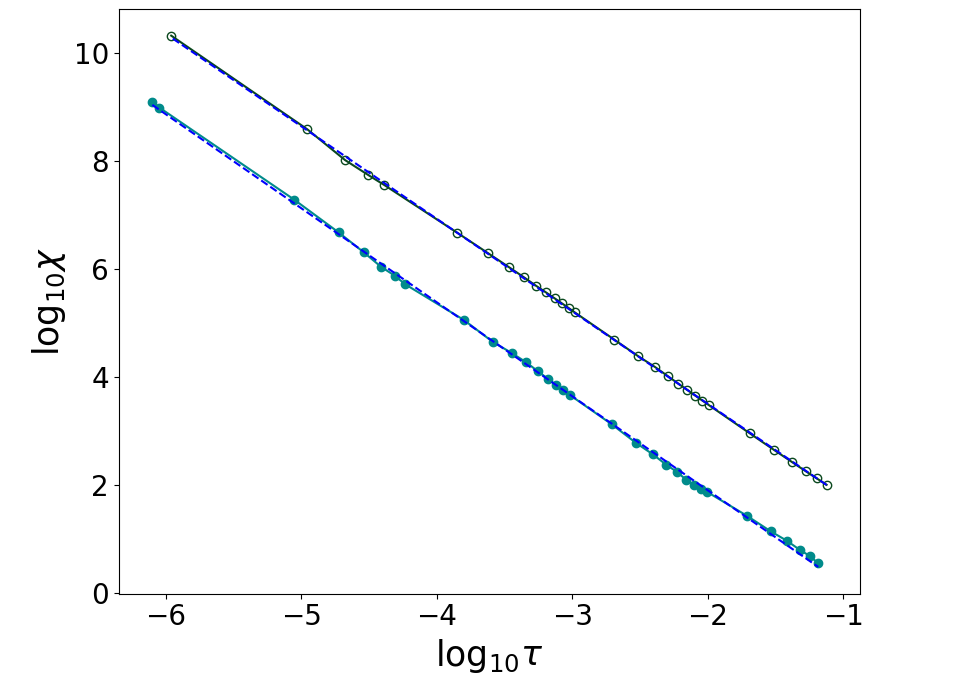}
\end{adjustbox}
\quad
\begin{minipage}[t]{0.05\columnwidth}
\subcaption{}
\end{minipage}
\begin{adjustbox}{minipage=0.88\columnwidth, valign=t}
\includegraphics[width=\columnwidth]{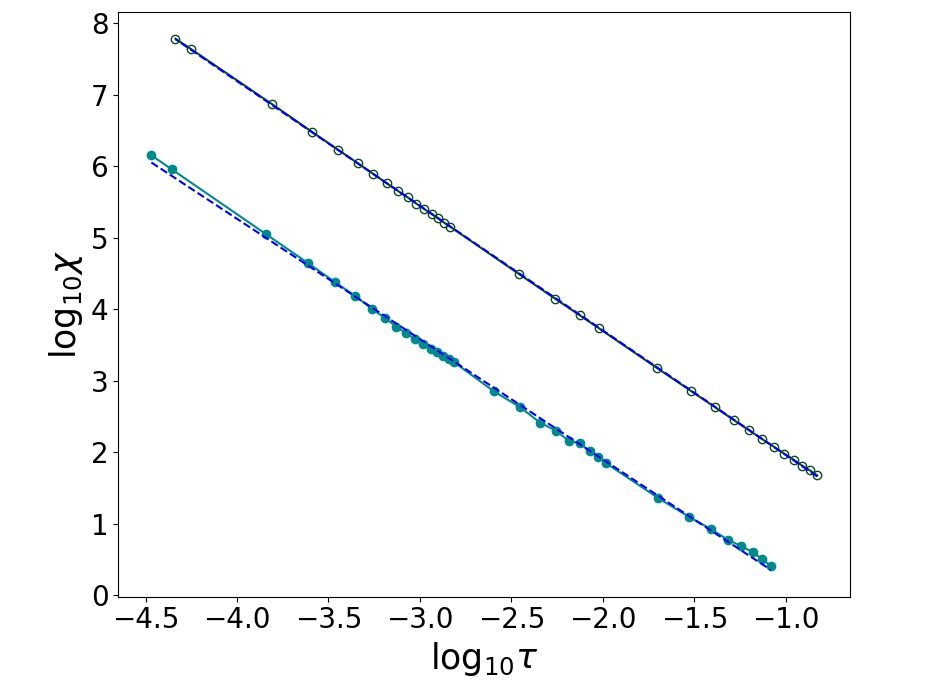}
\end{adjustbox}

%\subcaptionbox{}{\includegraphics[width=0.87\columnwidth]{susceptibility_log}}
%\subcaptionbox{}{\includegraphics[width=0.87\columnwidth]{anis10_susceptibility_log}} \\\vspace{-1mm}
%\subcaptionbox{}{\includegraphics[width=0.87\columnwidth]{correlation_comparison}}
%\subcaptionbox{}{\includegraphics[width=0.87\columnwidth]{correlation_comparison_anisotropic}}
%\\\vspace{-1mm}
%\subcaptionbox{}{\includegraphics[width=0.87\columnwidth]{criticalexp_delta_20}}
%\subcaptionbox{}{\includegraphics[width=0.87\columnwidth]{criticalexp_delta_20_anis10}}
%\\\vspace{-1mm}
%\subcaptionbox{}{\includegraphics[width=0.87\columnwidth]{amplitude_ratio_gamma}}
%\subcaptionbox{}{\includegraphics[width=0.87\columnwidth]{anis10_amplituderatio_gamma}}
\end{center}
\vspace{-4mm}
\caption{(color online) Left (right) column are data from the TRG flow with bond dimension $D=20$ on isotropic (anisotropy 10) Ising model in infinitesimal magnetic field $h=10^{-12}$. (a,b) Inverse temperature scan of susceptibility $\chi(\beta)$ showing the critical divergence at $\beta_C$ (red dashed line) slightly displaced from the true value $\beta_C^\text{Onsager}=1/T_C$ (blue dashed line). 
%We obtain the numerical critical temperature $\beta^D_c$ by looking at the plot of $\log \chi$ as a function of the temperature $\beta$. The $x$--axis in this plot is the dimensionless temperature $\beta/\beta^{\text{Onsager}}_c$ to clearly denote the error in $\beta^D_c$ relative to the analytical critical temperature, denoted by the navy dashed line obtained from Onsager's solution \cite{Onsager}. 
%The red dahsed-dotted line shows the location of $\beta^D_c$ relative to $\beta^{\text{Onsager}}_c$. 
\textit{Inset}: The plot of $\log \chi$ versus the iteration index of the TRG algorithm at $\beta_C$ clearly shows the expected finite-size scaling of the singularity, from which we extract $\gamma/\nu$; 
%shows the susceptibility does not reach infinity, indicating the effects of finite size-scaling.
(c,d) Critical and slightly off-critical correlations follow powerlaw decay out to scales determined by residual relevant perturbations, here corresponding to deviations around separations $= 2^{13}$;
(e,f) Non-linear critical magnetization displays a clear powerlaw with an associated critical exponent $\delta$;
(g,h) Off-critical susceptibility clearly showing scale invariant behavior assoicated with exponent $\gamma$ and also amplitude ratio $A\approx 45$ for the isotropic and $A\approx 48$ for the anisotropic, roughly consistent with the the known\cite{mccoyetal76amplitudes,delfino1998universal} exact value of $\approx 38$.}
\label{f:allisingplots}
\end{figure*}

 \section{results}
\label{sec:results}
\begin{figure*}
%\includegraphics[width=\linewidth]{YL_combined1}
%\caption{(color online)}
\begin{center}

\begin{minipage}[t]{0.05\columnwidth}
\subcaption{}
\end{minipage}
\begin{adjustbox}{minipage=0.95\columnwidth, valign=t}
\includegraphics[width=\columnwidth]{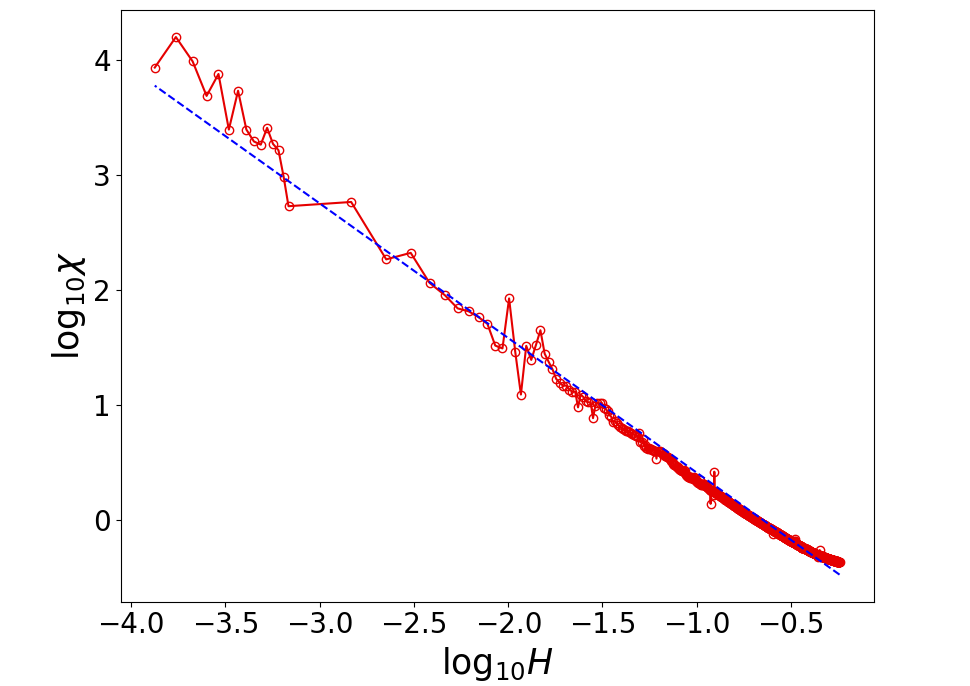}
\end{adjustbox}
\quad
\begin{minipage}[t]{0.05\columnwidth}
\subcaption{}
\end{minipage}
\begin{adjustbox}{minipage=0.95\columnwidth, valign=t}
\includegraphics[width=\columnwidth]{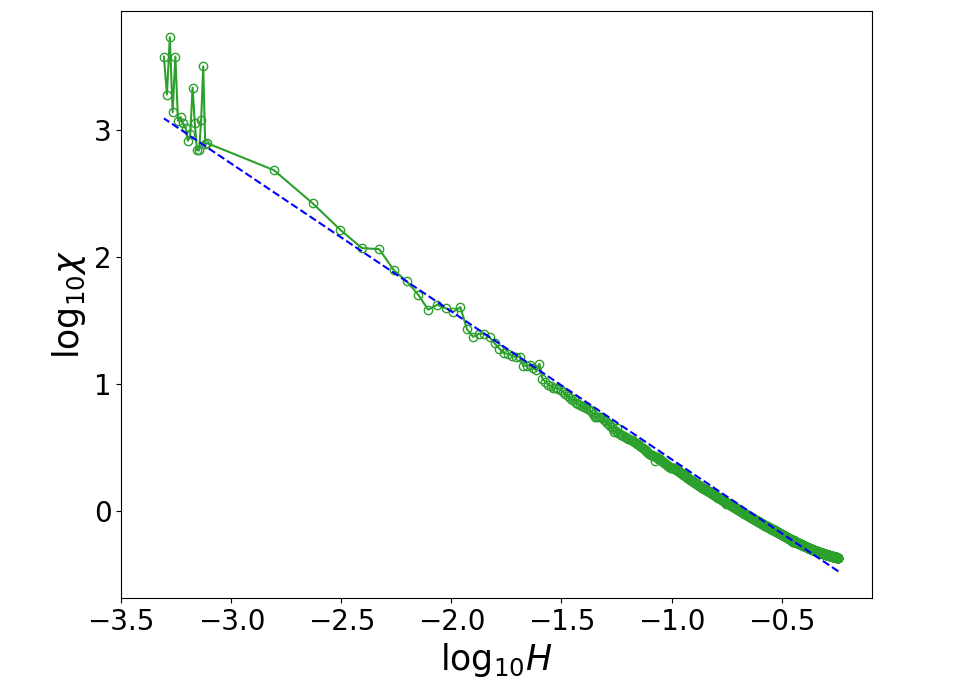}
\end{adjustbox}

\begin{minipage}[t]{0.05\columnwidth}
\subcaption{}
\end{minipage}
\begin{adjustbox}{minipage=0.95\columnwidth, valign=t}
\includegraphics[width=\columnwidth]{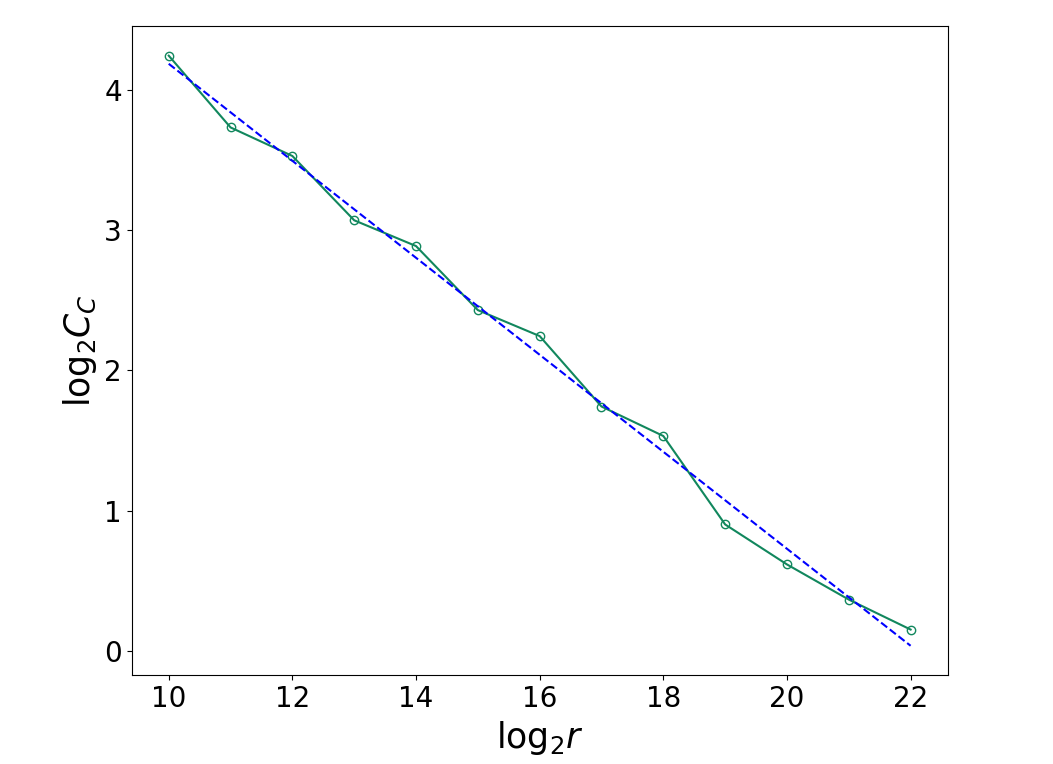}
\end{adjustbox}
\quad
\begin{minipage}[t]{0.05\columnwidth}
\subcaption{}
\end{minipage}
\begin{adjustbox}{minipage=0.95\columnwidth, valign=t}
\includegraphics[width=\columnwidth]{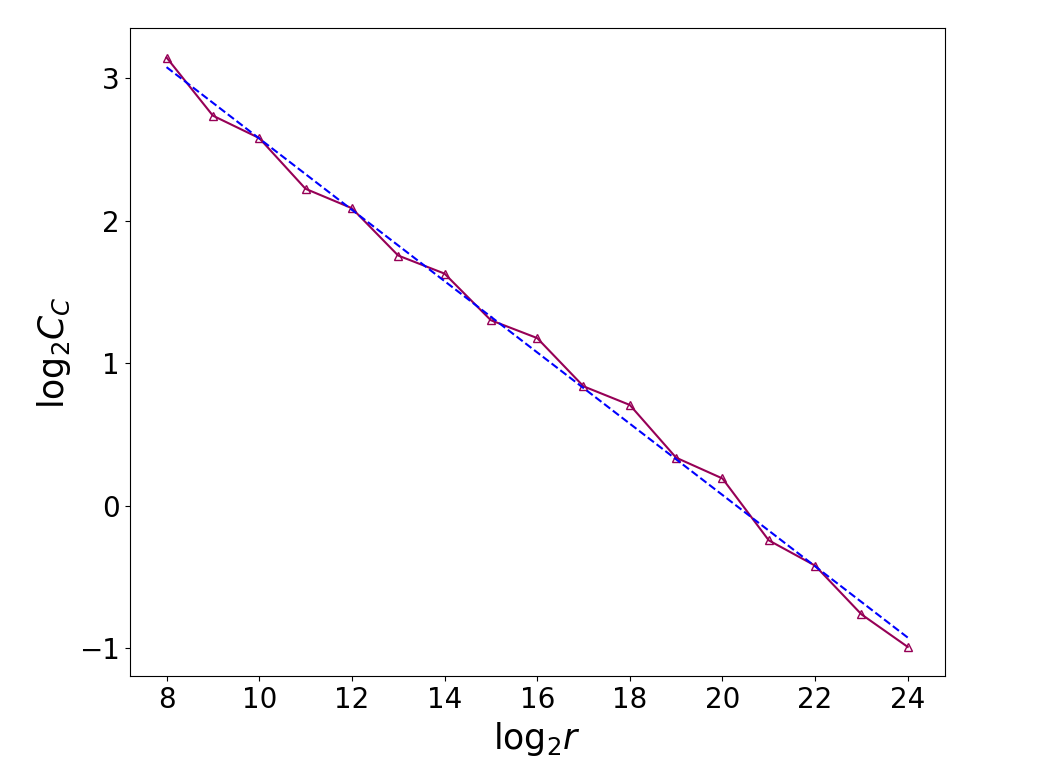}
\end{adjustbox}
%\subcaptionbox{}{\includegraphics[width=0.95\columnwidth]{YL_critexp_sigma16}}
%\subcaptionbox{}{\includegraphics[width=0.95\columnwidth]{YL_critexp_sigma20}}
%\\\vspace{-1mm}
%\subcaptionbox{}{\includegraphics[width=0.95\columnwidth]{YL_critexp_eta16}}
%\subcaptionbox{}{\includegraphics[width=0.95\columnwidth]{YL_critexp_eta20}}
\end{center}
\vspace{-3mm}
\caption{(color online) The top panel (row) shows the plots of $\log \chi$ as a function of $\log H$ exhibiting powerlaw behavior, where $H = h - ih_C$ for (a) $D = 16$; and (b) $D = 20$ from where we extract the critical exponent $\sigma$. In the bottom panel, we show the plots of $\log G(r,H)$ as a function of $\log r$ for (c) $D = 16$; and (d) $D = 20$. The decay for the connected correlator $G(r,H)$ is slow with critical exponent $\eta$. The critical exponents thus obtained are in good agreement with the analytical values $\sigma^{\text{Ana}} = -1/6$, and $\eta^{\text{Ana}}= -4/5$. See Sec. \ref{sec:resultsYL} for details.}
\label{f:YLplots}
\end{figure*}

This Section contains all the main results of the paper. % on universal properties of isotropic and anisotropoic Onsager and Yang-Lee critical points. 
For each bond dimension $D$ we perform a systematic scan of temperature and field (for the Ising-Onsager critical point) and complex field (for the Yang-Lee critical point) and collect the data into log-log plots (see Figs. \ref{f:allisingplots} and \ref{f:YLplots}) from which critical temperatures and critical exponents are extracted (and amplitude ratios for susceptibility) (see Tables \ref{t:isotropictable} and \ref{t:anisotropictable}).
For the Onsager-Ising critical point we focus on contrasting the results from isotropic and anisotropy 10 square lattice simulations (left and right columns in Fig. \ref{f:allisingplots}). The Yang-Lee critical point appears considerably more challenging for TRG but still exhibits a clear convergence trend in bond dimension.  We only consider quantities for which exact values are known, e.g. critical temperatures and exponents, and so for any such quantity X, we define a dimensionless error $\Delta X\equiv X^{\text{approx}}/X^{\text{exact}}-1$.  For critical (inverse) temperature $\beta_c$ we use a slightly more elaborate notation, with a superscript to denote the method used to extract the estimate of critical inverse temperature $\beta_c$  -- $\chi$ and $m$ for susceptibility and magnetization, respectively.

\subsection{The Onsager-Ising critical point}
\label{sec:resultsIsing}

Isotropic 2D Ising model was used as the default model for benchmarking most advances in tensor renormalization\cite{LevinNaveTRG,SRG_PRL,SRG_PRB,HOTRG,EvenblyTNR,LoopTNR}.  It has been appreciated, however, that this model is somehow non-generic, e.g. it has artificially small corrections to scaling and therefore may not be ideal for studying convergence properties of an approximate method.  However, much of past work focused on non-universal observables, e.g. the free energy and critical temperature. By contrast, we focus on universal quantitities and find (see Table \ref{t:isotropictable}) at best very preliminary signs of convergence in the limited range of bond dimension explored here.
%This does not appear to be the case for the set critical exponents to be inconclusive -- with accurate estimates emerging already at low bond dimension with no further signs of improvement for larger bond dimension and hence no basis for extrapolation to infinite bond dimension -- . 

The 2D Ising model in zero field is exactly solvable for arbitrary anisotropy, it exhibits identical universal scaling at the Onsager critical point regadless of anisotropy. %Equivalence of amplitude ratio $A = A_>/A_< $ computed at $D=20$ for the isotropic and anisotropic Ising models shows that these two models belong to the same \emph{universality class}.  
We do expect for a strongly anisotropic Ising model to exhibit a crossover from one to two dimensional behavior scaling which may make it somewhat challenging to capture for an approximate computational scheme or at the very least have a richer scaling function. Additionally, as we recently found \cite{Basuetal2021}, while the isotropic Ising critical point is a simple critical endpoint of a first order transition in the complex temperature plane, the anisotropic Ising critical point is a multicritical corner-point of a critical ordered phase, with modulated quasi-periodic order \cite{Basuetal2021}. This does not change any universal properties along the real temperature axis, as far as we know). Thus, the immediate proximity of a critical phase (albeit at complex temperature) is another reason to be interested in repeating the analysis of critical scaling to contrast the performance of TRG between isotropic and anisotropic Ising models. The results in Table \ref{t:anisotropictable} for anisotropy 10 were partly expected, i.e. larger deviations from exact values at smaller bond dimensions. 
Remarkably (and surprisingly), however, we also observe a clear convergence trend in universal exponents in bond dimension, see Fig. \ref{f:anisconvtrend}, in some cases to an order of magnitude better than for the isotropic model, e.g. for $\gamma$ and $\gamma/\nu$.  Empirically we observe $~1/D$ and $1/D^5$ convergence for pairs $\eta, \delta$ and $\gamma, \gamma/\nu$ of exponents, respectively.
We do not have any clear understanding of this result nor theory for what one might expect for the asymptotic convergence in bond dimension, as $1/D\to 0$ (see also Sec. \ref{sec:summary}).  Lastly, we note that both sets of computations reproduce one of the most unusual aspects of the Ising-Onsager critical point -- the unusually large amplitude ratio for the susceptibility (see bottom panels of Fig. \ref{f:allisingplots} and corresponding caption).

\subsection{The Yang-Lee critical point}
\label{sec:resultsYL}
First order transitions may be thought of as branch cuts in the (possibly enlarged) space of couplings. The seminal work by Yang and Lee details the trajectory of one such branch cut, corresponding to a ferromagnet in a field, $h$. Using the fundamental theorem of algebra they were able to establish explicitly that the familiar transition that takes place at $h=0$ in the ferromagnetic phase $(T<T_c)$ recedes to finite purely complex fields for $(T>T_c$), with corresponding branch points (critical endpoints) located at $h_c=\pm i |h_c|$.\footnote{The formalism of analytic continuation of couplings proved useful, e.g. in establishing the existence of essential (Griffiths) singularites in disorded magnets.} The Yang-Lee endpoints are novel, non-Onsager critical points. They are sometimes referred to\cite{FisherPRL78} as ``proto-critical'' points for having a single relevant perturbation, the complex magnetic field. They were identified with a complex scalar $\varphi^3$  field theory by Fisher\cite{FisherPRL78} and, in two dimensions,  the minimal model $\mathcal{M}(5,2)$ by Cardy\cite{CardyPRL85}.  Hence, the exact critical exponents $\sigma=-1/6$ and $\eta=-4/5$ are known. Importantly for what follows, the standard two-point scaling ansatz in the vicinity of the Yang-Lee critical point reads
\begin{equation}
C_C(r,\xi)=D(r/\xi) r^{d-2+\eta}
\end{equation}
with correlation length $\xi\sim |h-h_c|^{-\nu}$ and a \emph{singular} scaling function $D(x)=A \exp(-B x)/x^2$, where $A,B$ are constants. Importantly, the additional singularity in $D(x)$ implies a divergent \emph{amplitude} of the correlator $\sim \xi^2$.

Monte-Carlo techniques are powerless to access the Yang-Lee and other similarly exotic critical points, so the analysis historically relied on a combination of methods, including high temperature series \cite{griffiths69,Kurtze79}, analysis of partition function zeroes in finite lattices, culminating in matching to known conformal field theories\cite{CardyPRL85,CardyMussardo}.  Tensor network formulation was applied to this problem\cite{GarciaWeiYL} but did not consider two-point correlation function $C_C(r)$ nor susceptibility.  Our results for these two quantities are presented in Fig. \ref{f:YLplots}. %As we approach the Yang-Lee critical endpoints as was also observed by the authors in \cite{GarciaWeiYL} in their application of the HOTRG algorithm near the critical points. 
In spite of the noisy data, there is clear evidence of powerlaw dependences from which we extract $\sigma$ and $\eta$. The errors made for bond dimension $D=16$ and $20$ are, respectively, $\Delta_\sigma=-0.02, \Delta_\eta=0.18$ and $\Delta_\sigma=0.01, \Delta_\eta=0.06$. These preliminary results appear encouraging to expect a convergence trend as with the anisotropic Ising model.

\section{Summary and outlook}
\label{sec:summary}
To summarize, we computed several critical exponents of Ising and Yang-Lee critical points using TRG method to obtain and analyse correlation functions directly. We were especially interested to study convergence trends with bond dimensions and documented clear convergence in anisotropic Ising model and for the Yang-Lee critical point but not for the isotropic Ising model.  At the present we do not have a clear understanding of this observation and plan to explore it further. One candidate culprit is the onset of spurious fixed points of TRG near the critical point as has been previously documented \cite{TEFR,EvenblyTNR} for the isotropic model in zero field. It would be interesting to explore the convergence trends of other more sophisticated methods of tensor renormalization and also compare to the values of exponents obtained by other methods, e.g. from scaling dimensions using CFT ansatz for the finite size free energy.  Tensor renormalization techniques can also be fruitfully applied to statistical mechanics problems with non-positive weights as demonstrated here on the Yang-Lee problem. There are several such interesting problems in the literature, e.g. models with complex fugacity (similar to Yang-Lee), non-linear sigma models with topological terms, problems with gauge fields (and long-range interactions), quantum circuits with measurement\cite{Basuetal2021} and we expect to see numerical progress on these in the near future. Additionally, we expect tensor methods to make impact on problems with competing phases, possibly separated by weak first order transitions which often present severe difficulty to Monte-Carlo techniques. While generic three dimensional problems are still likely too difficult for an effective application of TRG, strongly anisotropic lattices or lattices with low local coordination may turn out more amenable.
\begin{acknowledgments}
We would like to thank Miles Stoudenmire, Tzu-Chieh Wei, Nikko Pomata, Frank Pollman, Chris Laumann, Aleix Bou Comas for many illuminating conversations. The Flatiron Institute is a division of the Simons Foundation.
\end{acknowledgments}

%\bibliographystyle{unsrt}
%\bibliography{criticalTRGbib}
%\bibliography{thesis}

\end{document}